\documentclass[twoside,twocolumn,10pt]{article}
\usepackage[english]{babel}
\usepackage[a4paper,left=1.5cm, right=1.5cm, top=1.785cm, bottom=2.0cm]{geometry}
\usepackage{authblk}

\usepackage{graphicx} 
\usepackage{amsmath}
\usepackage{hyperref}
\usepackage[version=3]{mhchem}
\usepackage{balance}
\usepackage{lastpage}
\usepackage{fancyhdr} 
\usepackage{fnpos} 
\usepackage[T1]{fontenc}
\usepackage[usenames,dvipsnames]{xcolor}
\usepackage{setspace}
\usepackage[compact]{titlesec}

\usepackage{droidsans} 

\usepackage[style=chem-acs, backend=bibtex]{biblatex}
\usepackage{upgreek}
\usepackage{siunitx}

\usepackage[format=plain,justification=justified,singlelinecheck=false,font={small,sf},labelfont=bf,labelsep=space]{caption}

\date{}
\addto{\captionsenglish}{}

\title{Competing excitonic couplings as origin of mimicked phase transitions in zinc-phthalocyanine single crystals}

\author[1]{Lisa Schraut-May}
\author[1]{Sebastian Hammer}
\author[2]{Luca Nils Philipp} 
\author[3]{Krzysztof Radacki}
\author[4]{Gabriele Tauscher}
\author[1]{Helena Hollstein}
\author[1]{Kilian Strauß}
\author[5]{Martin Kamp}
\author[4]{Heinrich Schwoerer}
\author[3]{Holger Braunschweig}
\author[2]{Roland Mitric}
\author[1,6]{Jens Pflaum}

\affil[1]{\small \textit{Experimental Physics VI, Julius-Maximilian University Würzburg, Am Hubland, 97074 Würzburg, Germany. E-mail: lisa.schraut-may@uni-wuerzburg.de, sebastian.hammer@uni-wuerzburg.de, jens.pflaum@uni-wuerzburg.de}}
\affil[2]{\small \textit{Chair of Theoretical Chemistry, Julius-Maximilian University Würzburg, Am Hubland, 97074 Würzburg, Germany.}}
\affil[3]{\small \textit{Institute of Inorganic Chemistry, Julius-Maximilian University Würzburg, Am Hubland, 97074 Würzburg, Germany.}}
\affil[4]{\small \textit{Max Planck Institute for the Structure and Dynamics of Matter, Luruper Chaussee 149, 22761 Hamburg, Germany.}}
\affil[5]{\small \textit{Institute of Physics and Röntgen Center for Complex Material Systems, Julius-Maximilian University Würzburg, Am Hubland, 97074 Würzburg, Germany.}}
\affil[6]{\small \textit{Center for Applied Energy Research, 97074 Würzburg, Germany.}}

\begin{document}

\twocolumn[
\begin{@twocolumnfalse}
\maketitle
\begin{abstract}
\sffamily{The optical properties of molecular crystals are largely determined by the excitonic coupling of neighboring molecules.
This coupling is extremely sensitive to the arrangement of adjacent molecular units, as their electronic interaction is defined by the relative orientation of the individual transition dipole moments and their wave function overlap.
Hence, the optical properties, such as transmission, absorption, and emission, are usually highly anisotropic and good indicators of structural changes during the variation of intensive thermodynamic parameters like temperature or pressure.
Here, we discuss the peculiar though archetypical case of $\upbeta$-phase zinc-phthalocyanine: In single crystalline specimen, we report a sudden change of spectral emission with temperature from a broad, unpolarized Frenkel-exciton type luminescence to a narrow, highly polarized superradiance-like fluorescence below 80\,K. 
Surprisingly, we find that there is no sign of a discrete structural phase transition in this temperature regime.
To understand this apparent contradiction, we perform polarization-, temperature- and time-dependent photoluminescence measurements along different crystallographic directions to fully map the emission characteristics of the crystal-exciton.
By means of ab-initio calculations on a density functional theory level we conclude that our observations are consistent with a dimer exciton model when considering thermalized electronic states.
As such, our study presents a representative case study on a well-established molecular material class demonstrating that caution is advised when attributing discrete changes in electronic observables to a structural phase transition. 
As we show for zinc-phthalocyanine in its  $\upbeta$-phase modification, a material that has received strong attention for its use in various optical and opto-electronic applications, slowly varying excitonic couplings and thermal redistribution of excitations can mimic the same signatures attributed to a structural phase transition.}
\end{abstract}
\vspace{3mm}
\end{@twocolumnfalse}
]





\section{Introduction}

Molecular aggregates are comprised of extended, spatially anisotropic constituents.
Strong ground state associated dipole moment forces in combination with superimposed weak dispersion forces lead to crystallization of molecular single crystals in low symmetry point groups, often being monoclinic or even triclinic in nature.\cite{Schwoerer.2007}
The electronic interaction in the aggregate depends on the relative orientation of adjacent molecules and is therefore highly anisotropic as well.\cite{Hestand.2018} 
This gives rise to complex photophysical behavior that depends on the aggregate’s underlying crystal structure and is sensitive to disruptions of the local molecular geometry. 

A common description of the optical properties is given in terms of exciton theory. 
Based on the work of Kasha and coworkers,\cite{Kasha.1950,Kasha.1965} the excitonic couplings are often described by dipolar Coulomb interactions between the transition dipole moments (TDM) of adjacent molecules.
In the case of one-dimensional chains this leads to the distinction of two limiting cases, the H- and J-aggregate.\cite{McRae.1958}
The treatment of the neat electronic interactions can be expanded to include coupling to the nuclear coordinates to take intra- and inter-molecular vibrations into account. 
This results in the exciton-polaron model, which provides a much more comprehensive picture of the photo-physics of molecular aggregates, especially regarding its dependence on temperature and dynamic and static disorder.\cite{Hestand.2018,Spano.2010}
These approaches have proven successful in describing the optical properties of various molecular materials and concomitant crystal structures, from one-dimensional linear\cite{Hestand.2018} and helical chains\cite{Rodel.2025} to the more sophisticated cases of two-dimensional brick-wall\cite{Eisfeld.2017} or herringbone aggregates, \cite{Spano.2004,Spano.2005} as well as excimer formation.\cite{Bialas.2022}
From this it is quite clear that the optical properties of a molecular aggregate can be used as indicator for the respective underlying polymorph.\cite{Yu.2010,Gierschner.2021,Nogueira.2020} 
Especially the spectral photoluminescence has been used as a non-destructive probe to study structural phase transitions of molecular crystals in-situ.\cite{Hammer.2019,Ge.2018,Xu.2016} 
Here, the knowledge of the underlying crystal structure as well as the specific orientation of the molecules in the probed crystallographic facets is extremely important for the correct interpretation of the observed photoluminescence signals and their spectral distribution. 

The following study is motivated by the photoluminescence of $\upbeta$-phase zinc-phthalocyanine ($\upbeta$-ZnPc) single crystals (for molecular and crystal structure \textit{c.f.} Figure~\ref{fgr:1-PL}~a) and~b)).
ZnPc is a well-established organic material, studied for its optical properties and utilization in opto-electronic applications, like organic photovoltaic,\cite{Senthilarasu.2003, Riede.2011, Brendel.2015, Jungbluth.2023} light emitting diodes\cite{Hammer.2019} or even novel concepts such as plasmonic light emitting antennas. \cite{Grimm.2022, Rodel.2022}
Despite the broad interest in this material, its optical behavior over a wide temperature range or specifically at low temperatures was, to our knowledge, never studied in detail before.
In Figure~\ref{fgr:1-PL}~c) the evolution of the photoluminescence is shown as a function of temperature. 
For temperatures above 150~K, the spectra represent a typical Frenkel-type exciton emission as shown exemplary in Figure~\ref{fgr:1-PL}~d) for 290~K. 
Cooling down further, around 100~K, the main emission peak at 1.6~eV disappears almost completely (\textit{c.f.} center of Figure~\ref{fgr:1-PL}~d)) and reappears as a narrow emission peak at low temperatures as depicted in the lower panel of Figure~\ref{fgr:1-PL}~d). 
The emergence of such narrow emission peaks of high intensity in the lowest emission component is known from low-temperature J-aggregates and commonly termed superradiance. 
To understand this behavior, we present a comprehensive study of temperature-, polarization-, and time-dependent photoluminescence on different crystallographic facets, consistently describing the complex luminescence behavior of $\upbeta$-ZnPc single crystals.
Together with structural characterization and complementary theoretical evaluation of the corresponding excitonic coupling between neighboring molecules via time-dependent density functional theory (TD-DFT), we clearly demonstrate that no underlying structural phase transition is responsible for this drastic change in the emission spectra with temperature. 
Instead, the observed behavior can be consistently described by an elementary aggregate-exciton model if thermal population of dark and emissive states is fully taken into account. 

\section{Temperature dependent photoluminescence}
\subsection{Steady-state photoluminescence}

\begin{figure*}[htb!]
 \centering
 \includegraphics{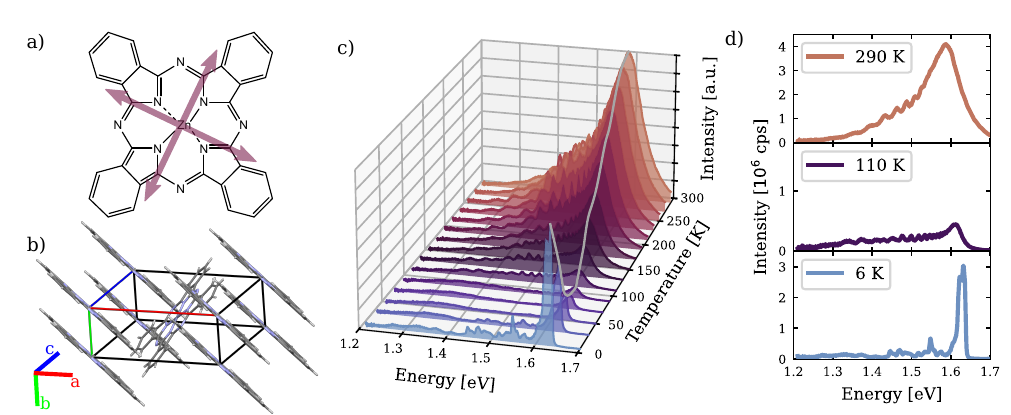}
 \caption{a) Molecular structure of ZnPc with purple arrows indicating the TDM orientation of the $\text{S}_0 \rightarrow \text{S}_1$ and $\text{S}_0 \rightarrow\text{S}_2$ calculated via TD-DFT. b) Unit cell of $\upbeta$-phase ZnPc with herringbone-like packing. c) Temperature dependent PL spectra measured on the ZnPc $(\bar{1}01)$ crystal facet with a gray line as guide to the eye indicating the evolution of spectral maximum with temperature. The spectra exhibit Fabry-Pérot oscillations at lower energies being indicative for the high crystal quality. d) Luminescence spectra measured at exemplary temperatures of 290\,K, 110\,K and 6\,K.}
 \label{fgr:1-PL}
\end{figure*}

We first turn to the temperature dependent fluorescence spectra of the investigated $\upbeta$-ZnPc and their detailed description. 
The measurements were conducted via micro-photoluminescence on  $\upbeta$-ZnPc single crystals grown via horizontal vapor deposition (\textit{c.f.} Methods section for details).
The emission spectra are shown in Figure~\ref{fgr:1-PL}~c) for the temperature range from 6\,K to 300\,K.
The probed crystallographic facet was identified by X-ray diffraction as the $(\bar{1}01)$ bulk crystal facet.
The polarization direction of the depicted PL corresponds to the maximum emission, as discussed below. The emission polarization is perpendicular to the crystallographic b-axis which is the short herringbone stacking direction and macroscopically defines the crystals' long needle axis (\textit{c.f.} Figure~S1~d) to f) in SI).

It should be noted that the distinct intensity oscillations superimposing the spectra  at low energies are Fabry-Pérot oscillations caused by internal reflections at the outer crystal surfaces.
In case of highly parallel aligned crystal surfaces, the crystal acts as a resonator with selective standing wave modes.
This results in a modulation of PL intensity as one would find for an interference filter.\cite{Weber.1995} 
The occurrence of this interference phenomenon requires high-quality crystals with smooth and almost perfectly aligned parallel surfaces and is considered a measure for the high sample quality. 

The variation of the sample temperature reveals a complex emission behavior which can quantitatively be divided into two regimes:
The emission below 100\,K, depicted in the bottom panel of Figure~\ref{fgr:1-PL}~d), is characterized by a single prominent, narrow emission line at 1.63\,eV (FWHM of 20\,meV) that vanishes with increasing temperatures (center panel of Figure~\ref{fgr:1-PL}~d)).
For temperatures above 100\,K, a broad emission emerges which increases in intensity with rising temperature reaching a FWHM of 80\,meV for the main peak at 1.59\,eV at 290\,K as depicted in the upper panel of Figure~\ref{fgr:1-PL}~d).
 
The room temperature photoluminescence is consistent with emission spectra reported for ZnPc $\upbeta$-phase thin films.\cite{Hammer.2019, Kato.2020, Doctor.2021}
Therefore, we relate the observed emission to a Frenkel-type excitonic state with a main transition around 1.59\,eV and a broad vibronic progression towards lower energies. 
The two lowest excited singlet states of the ZnPc monomer are only almost degenerate, as the D$_{4\text{h}}$ symmetry of the molecule is slightly distorted in the crystal.
The orientations of the related TDMs are illustrated in Figure~\ref{fgr:1-PL}~a). However, as the corresponding energy splitting is of the order of only a few tens of millielectronvolts,\cite{Feng.2020} this effect is masked by the broad emission peak at higher temperatures. 

In contrast, we attribute the very narrow emission peak in the cryogenic temperature regime to a J-aggregate emission. 
Here, the coherent coupling between several neighboring molecules enhances the transition between the vibrational ground state of the excited state and the vibrational ground state of the electronic ground state (notated 0-0), compared to transitions to higher vibrational modes in the electronic ground state (notated 0-n).\cite{Hestand.2018}

\begin{figure}[htb]
\centering
  \includegraphics[]{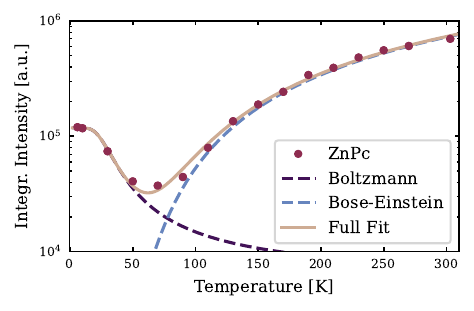}
  \caption{Integrated intensity of the spectra displayed in Figure~1~c) as function of temperature showing a Boltzmann activated decay at low temperatures and an increase in emission at higher temperatures following the Bose-Einstein population of phonons.}
  \label{fgr:2-PL-Fit}
\end{figure}

As demonstrated in several experimental and theoretical studies on other herringbone aggregates,\cite{Hestand.2018,Spano.2004,Spano.2005,Tanaka.2016} the temperature dependent characteristic of the ZnPc fluorescence cannot consistently be explained by a simple J- or H-aggregate emission alone.
Rather a combination of both is needed to consider the differing molecular packing along different crystal directions.
The integrated intensity shown in Figure~\ref{fgr:2-PL-Fit}, can be described by the superposition of two different temperature-dependent processes.
An overall decrease of the emission with rising temperature is commonly observed for J-aggregates and is related to the loss of coherent coupling by scattering with thermally activated inter- and intra-molecular vibrations.
Furthermore, thermal activation of energetically higher dark states or non-radiative pathways can lead to an overall decrease in emission intensity for J-aggregates with rising temperature. \cite{Potma.1998, Hestand.2018}
In contrast, for H-aggregates the opposite behavior appears.
For a momentum allowed radiative decay process to occur, the exciton needs to scatter with a phononic or vibrational excitation. The emission of H-aggregates increases with rising temperature, as the population of high-momentum phonons, and therefore the rate of scattering processes, increases.\cite{Spano.2010,Hestand.2018}

This suggests that the two types of aggregate emission dominate different temperature regimes in ZnPc:
The high temperature regime is dominated by H-aggregate luminescence, whereas a J-aggregate emission dominates at low temperatures. 
The high temperature emission can be modeled to first order by a Bose-Einstein phonon population with a single phonon energy of $E_\textrm{ph}=(29.3\pm2.1)$\,meV.
We interpret this in an H-aggregate emission picture: A low energy vibrational mode mediates the thermal population of the Brillouin zone center. 
From there the transition to the electronic and vibrational ground state can take place, effectively increasing the spectral contribution at 1.6\,eV compared to the vibronic tail below.
Furthermore, a second driving factor for the increase in emission with rising temperature is given by the thermal population of additional emissive states leading to an overall intensity increase as will be discussed further below. 
Below 100\,K, the intensity decreases with temperature following a Boltzmann activated quenching process with an activation energy of $E_\textrm{A}=(9.5\pm2.1)$\,meV.
This fits well into the picture of a J-aggregate emission being depopulated via thermal excitation of dark states or non-radiative pathways.

The temperature dependent change in the emission characteristic is furthermore accompanied by a gradual shift of the main peak with rising temperature towards lower energies of about 40\,meV over the whole temperature range.
We attribute this shift partly to self-absorption in the bulk crystal, as an overall peak broadening together with self-absorption can lead to an apparent red-shift in the luminescence signal,\cite{Irkhin.2012} as shown for emission of our ZnPc $(\bar{1}01)$ facet at higher temperatures (\textit{c.f.} Section S2 in SI).

Upon the transition from J- to H-aggregate emission with increasing temperature, the observed dramatic change in luminescence behavior intuitively suggest some form of accompanying change of the underlying molecular packing.
Interestingly enough, we do not find any evidence of a structural phase transition, neither in the crossover temperature regime, nor well below the emergence of the low temperature emission.
We rather find a gradual, though highly anisotropic change in the unit cell parameters (\textit{c.f.} Section S3 in SI).  

\bigskip

\subsection{Spatial anisotropy of polarization dependent photoluminescence}

\begin{figure}[htb]
\centering
  \includegraphics[]{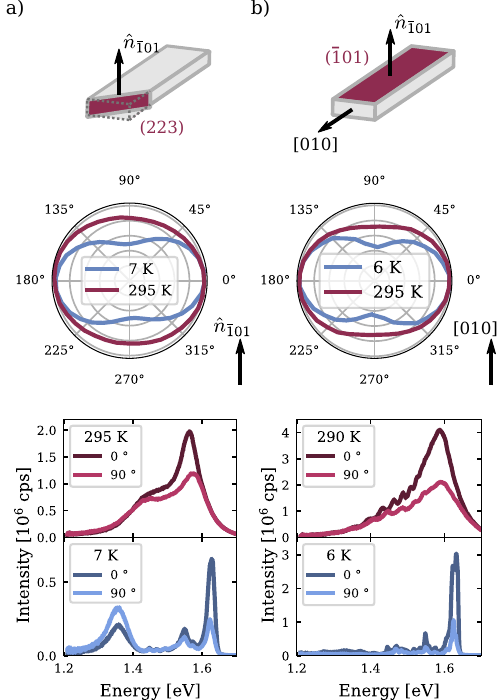}
  \caption{Comparison of the integrated intensity as a function of the polarization angle measured in respect to the a) microtome cut $(223)$ plane and (b) naturally occurring $(\bar{1}01)$ facet,  illustrated by the schematic above. For the $(223)$ facet at 7\,K only the spectral region above 1.45\,eV is included for integration. 90\,$^\circ$ polarization refers to the normal vector $\hat{n}_{\bar{1}01}$ of $(\bar{1}01)$ plane and the $[010]$ crystallographic direction, respectively. At the bottom, representative spectra for the polarization directions of highest (0\,$^\circ$) and lowest (90\,$^\circ$) intensity are shown.}
  \label{fgr:3-Pol}
\end{figure}

For two-dimensional herringbone aggregates a significant dependence of the emission on different crystallographic directions is reported both theoretically and experimentally due to the directionality of the excitonic coupling of the involved molecules.\cite{Spano.2004,Spano.2005,Meinardi.2003}
To enable the optical characterization of a different crystallographic direction besides the naturally occurring and therefore easily accessible $(\bar{1}01)$ facet, cross-sections of the needle shaped ZnPc single crystals roughly perpendicular to their long needle axis were prepared by ultra-microtome sectioning. 
With selected area electron diffraction (SAED) we identify the resulting crystal facet as the $(223)$ crystal plane \textit{(c.f.} Figure~S4 in SI).

In Figure~\ref{fgr:3-Pol} we present a comprehensive data set of polarization dependent photoluminescence measurements on the $(\bar{1}01)$~(a) and $(223)$ facets~(b) for low (blue) and high (purple) temperature, respectively. 
The top part illustrates the orientation of the probed crystal facets. 
The center plots display the integrated intensity as a function of emission polarization angle with respect to a reference crystal direction (90\,$^\circ$) and the emission maximum (0\,$^\circ$).
For the $(223)$ plane in a) the reference direction is the normal vector $\hat{n}_{\bar{1}01}$ of $(\bar{1}01)$ plane (note that this direction is not equal to the crystallographic $[\bar{1}01]$ direction due to the monoclinic unit cell of ZnPc, \textit{c.f.} Figure S1~c) in the SI) and for the $(\bar{1}01)$ facet in b) the crystallographic $[010]$ direction along the long needle axis of the single crystal. 
At the bottom, exemplary spectra for low and high temperatures, as well as maximum (0\,$^\circ$ polarization) and minimum (90\,$^\circ$ polarization) emission intensity are shown. 

The emission of the naturally occurring $(\bar{1}01)$ facet (Figure~\ref{fgr:3-Pol}~b)) is strongly polarized at low temperatures.
The emission maximum lies perpendicular to the face-to-face herringbone stacking direction, \textit{i.e.} the crystallographic $[010]$-axis (\textit{c.f.}  right panel of Figure~\ref{fgr:5-tdm} b)). 
While the general polarization direction is maintained with rising temperature, the degree of polarization decreases, meaning the polarization becomes increasingly isotropic. 
For the emission of the $(223)$ plane, a similar behavior can be observed with a preferred polarization parallel to the $(\bar{1}01)$ plane and an even lower degree of polarization at elevated temperatures. 

This data set gives us unique insights into the directionality of the photophysical behavior, which is further emphasized by the emission spectra shown at the bottom of Figure~\ref{fgr:3-Pol}.
The intensity at low temperatures almost vanishes for the emission polarization angles of 0\,$^\circ$ for both facets, respectively. 
Additionally, the high-temperature spectra measured on both facets, most evident for the $(223)$-facet, show  a polarization dependent energetic shift, while no such dependence is visible at low temperatures.
This indicates that the broad room temperature emission likely encompasses several thermally populated emissive states differing in their polarization directionality.
A rather striking feature in the fluorescence of the microtome sections is the additional broad and featureless peak around 1.35\,eV.
Upon closer inspection, this feature is also present in the spectra of the naturally occurring $(\bar{1}01)$ facet, but with an extremely low intensity (\textit{c.f.} Figure~S5 in SI). 
This peak resembles the emission of the ZnPc $\upalpha$-phase stemming from an excited dimer, so-called excimer state, leading to a broad and red shifted emission.\cite{Bala.2009, Hammer.2024} 
Our extensive X-ray studies revealed pure $\upbeta$-phase molecular arrangement in the single crystals, thus we can rule out the presence of any $\upalpha$-phase admixture.
The excimer state has, to our knowledge, not yet been reported to coexist in the $\upbeta$-phase of ZnPc single crystals and
emphasizes that due to the strong directional dependence of the fluorescence, especially in low-symmetry systems as the one investigated in this study, not all emissive states can be probed by examining only one crystal facet.

\subsection{Lifetime analysis}

\begin{figure}[htb]
\centering
  \includegraphics[]{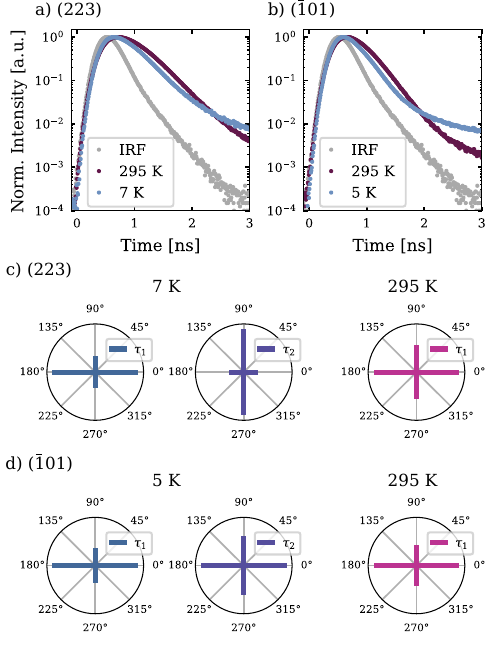}
  \caption{Fluorescence decay of the main emission peak around 1.6\,eV at 5\,K or 7\,K and 295\,K for the a) $(223)$ and b) $(\bar{1}01)$ facet with polarization angles of 0\,$^\circ$, meaning parallel to the $(\bar{1}01)$ plane and $[010]$-axis, respectively. The instrument response function (IRF) of the setup is shown in gray for comparison. Normalized amplitudes of the one (two) main components with lifetimes $\tau_1$ (and $\tau_2$) at 295\,K (5 or 7\,K) for both measured polarizations for the c) $(223)$ and d) $(\bar{1}01)$ facets. Polarization angles are defined analogously to Figure~\ref{fgr:3-Pol}.}
  \label{fgr:4-Lifetime}
\end{figure}

To fully disentangle the spectral components, we extend our steady-state study to time-dependent measurements of the fluorescence signal using time-resolved emission spectroscopy (TRES) for different emission polarizations and temperatures.
The measurements were conducted for the polarization directions of the highest and lowest steady-state emission intensity of the two crystallographic facets, respectively.
As a first step, we performed a global analysis (GA).
Using this first analysis, we assign the main electronic transitions to the region around 1.6\,eV for both facets, and the low energy tail as their accompanying vibronic progression and the excimer component at 1.35\,eV.
The full details of this assignment are discussed in the SI, Section S6.1.
Figure~\ref{fgr:4-Lifetime}~a) and b) depict the fluorescence decay of the spectral region around 1.6\,eV at highest and lowest measured temperatures for the $(223)$ and $(\bar{1}01)$ facet with polarization angles of 0\,$^\circ$, meaning parallel to the $(\bar{1}01)$ plane and $[010]$-axis, respectively
A change in emission lifetime between the two temperatures is obvious.
It is evident, that the radiative decay measured for the $(223)$ facet is overall slower, compared to that observed for the $(\bar{1}01)$ facet. 
We attribute this to a change in fluorescence quantum yield along different crystallographic directions due to ZnPc's strong anisotropy in both its optical and structural properties, influencing the experimentally accessible lifetime e.g. by non-radiative decay pathways.

The following temperature dependent behavior is derived for both facets and polarizations by subsequent multi-exponential fitting of the main feature of the spectrum around 1.6\,eV (see Section S6.2 in SI for details). 
The radiative lifetimes derived for the $(223)$ and $(\bar{1}01)$ facets with polarization of the emission being parallel to the $(\bar{1}01)$ plane and perpendicular to the $[010]$ direction, respectively, are listed in Table~\ref{tbl:lifetime}.
At room temperature, the spectrum is comprised of a single spectral component with a lifetime of around 250\,ps (360\,ps) measured for the $(\bar{1}01)$ facet ($(223)$ facet). 
Towards cryogenic temperatures, this fast emissive state decaying in the ps-range ($\tau_1$) decreases in lifetime to a minimum of 168\,ps (283\,ps) for the $(\bar{1}01)$ facet ($(223)$ facet). 
Furthermore, a second contribution ($\tau_2$) with approximately two orders of magnitude less intensity and a lifetime in the ns-range must be taken into account to fully describe the time dependence of the emission. 
From the relative intensities of these perpendicular polarized emissive components, one can deduce their directionality, though this is not an exhaustive polarization characterization. 
Their normalized intensities are depicted in Figure~\ref{fgr:4-Lifetime}~c) and d). 
The single component at high temperatures describes the entire spectrum. 
It is polarized perpendicular to the $[010]$ direction and along the $(\bar{1}01)$ plane, and thus follows the same directionality seen in the steady-state spectra. 

\begin{table}[b]
\small
  \caption{\ Fluorescence lifetimes $\tau_1$ and $\tau_2$ of the main peak in the spectral region around  1.6 eV derived by multi-exponential fitting. The values correspond to the $(\bar{1}01)$ and $(223)$ facet with polarization perpendicular to the $[010]$ direction and parallel to the $(\bar{1}01)$ plane, respectively.}
  \label{tbl:lifetime}
  \centering
  \begin{tabular*}{0.45\textwidth}{@{\extracolsep{\fill}}lllll}
    \hline
     & $(223)$ & & $(\bar{1}01)$ &  \\
    T/K & $\tau_1$/ps & $\tau_2$/ns & $\tau_1$/ps & $\tau_2$/ns \\
    \hline
    5 & 283 $\pm$ 8 & 2.08 $\pm$ 0.03 & 168 $\pm$ 8 & 2.14 $\pm$ 0.01  \\
    80 & 328 $\pm$ 8 & 2.27 $\pm$ 0.03 & 227 $\pm$ 8 & 2.21 $\pm$ 0.09\\
    200 & 303 $\pm$ 8 & - & 251 $\pm$ 8 & - \\
    295 & 360 $\pm$ 8 & - & 245 $\pm$ 8 & - \\
    \hline
  \end{tabular*}
\end{table}

On the other hand, the two low temperature components differ not only in their respective lifetimes, but also in their directionality. The fast component with $\tau_1$ closely follows the polarization dependence of the full steady-state measurement.
It shows a higher intensity perpendicular to the $[010]$ axis for the natural $(\bar{1}01)$ facet, and parallel to the $(\bar{1}01)$ plane for the microtome-cut section.
Surprisingly, the slow component differs in its directionality. For the $(\bar{1}01)$ facet its intensity is almost isotropic, while it is polarized perpendicular to the $(\bar{1}01)$ plane for the microtome sliced $(223)$ facet.

\section{Excitonic coupling of neighboring molecules}

To theoretically describe the microscopic processes, that determine the photophyics of the ZnPc single crystals, we carried out \textit{ab-initio} time-dependent density functional theory (TD-DFT) simulations of the excited state properties of different ZnPc dimer configurations contained in the crystal structure. 
For the ZnPc monomer, the calculations show that the two energetically lowest electronically excited states, $\text{S}_1$ and $\text{S}_2$, are nearly degenerate and have non-vanishing TDMs. 
Furthermore, the next energetically higher excited state, $\text{S}_3$, is almost 1.5\,eV higher in energy than $\text{S}_1$ and $\text{S}_2$.
Thus, the emissive states of the aggregates will be mostly comprised of the monomeric $\text{S}_1$ and $\text{S}_2$ states. 
The orientation of the TDMs of the $\text{S}_1$ and $\text{S}_2$ of the ZnPc monomer are depicted in Figure~\ref{fgr:1-PL} a) and are in agreement with previous literature.\cite{Zhang.2016b}

\begin{figure}[tb]
\centering
  \includegraphics[]{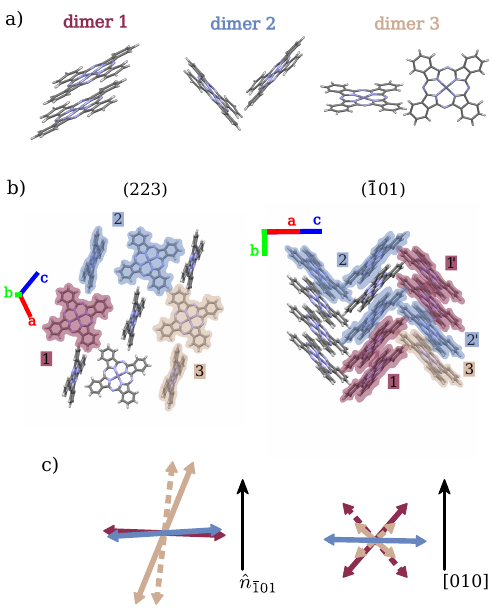}
  \caption{a) Depiction of the neighboring molecules representing the three investigated dimer configurations. b) Arrangement of the dimers in respect to the two measured crystallographic facets, the $(223)$ (left) and $(\bar{1}01)$ facet (right). Exemplary rotated versions of Dimers 1 and 2 are also shown for the $(\bar{1}01)$ facet and labeled Dimers 1’ and 2’ (see main text). c) Projection of the resulting TDMs of the energetically lowest transitions with non-vanishing moment onto the respective crystallographic planes above. The depicted length of the arrows therefore qualitatively correspond to the dipole strength. $\text{S}_3$ of Dimer 1 is indicated by a purple arrow, $\text{S}_1$ for Dimers 2 and 3 in blue and yellow, respectively. Additionally, the TDM of the rotated dimers are shown by dashed arrows, where the different orientations result in substantial changes in TDM orientation.}
  \label{fgr:5-tdm}
\end{figure}

A common approach for understanding the optical properties of aggregated monomers is to describe the aggregate by means of dimer-units of neighboring molecules. 
The reason for this is, that short-range coupling is the dominant contribution to the excitonic interactions and hence, many aggregates can be consistently described by a full quantum chemical treatment of the dimer units they contain.\cite{Engels.2017,Hestand.2015,Kasha.1965} 
We identified the three different ZnPc dimer configurations that make up the $\upbeta$-phase crystal structure as shown in Figure~\ref{fgr:5-tdm}:
Along the crystallographic b-axis, ZnPc monomers are $\uppi$-stacked in chains (Dimer 1). 
Monomers in adjacent chains have next neighbors either within the $(\bar{1}01)$ plane (Dimer 2) or along the $[\bar{1}01]$ out-of-plane direction (Dimer 3). 
Due to the inversion symmetry of the P$2_1$/n unit cell, each dimer has a mirror image noted with a prime in Figure~\ref{fgr:5-tdm}~b).

In Dimer 1 the molecules are stacked parallel in a face-to-face manner. 
Since the TDMs of the two lowest lying excited states of a ZnPc monomer lie within the plane of the molecule, Dimer 1 is likely a H-type aggregate.
The coupling between neighboring stacks (Dimers 2 and 3 in Figure~\ref{fgr:5-tdm}~a) center and right) on the other hand, is of J-type nature. 
More precisely, due to the corresponding TDMs of the monomers not being parallel to each other, of so-called X-aggregate type.\cite{Ma.2021,Lijina.2023}

For Dimer 1, meaning the coupling within the molecular chains, the calculations show that the TDMs of the two energetically lowest excited states vanish.
This corroborates our first crude categorization as an H-type aggregate by dipole orientation. 
Here however, the full dimer wave function is considered, also including coupling by wave function overlap.
According to Kasha’s rule, fluorescence occurs from the energetically lowest electronically excited state, unless the energy difference to higher-lying states is small enough to allow for thermal population of the higher excited state.
Therefore, Dimer 1 is not expected to contribute to the emission of the single crystal at low temperatures. 
Only at elevated temperatures Dimer 1 can contribute to the emission via thermal population of its higher excited states, particularly its $\text{S}_3$ state. 
The simulations reveal that the coupling in Dimers 2 and 3 is indeed of the J-aggregate type. 
The energetically lowest electronically excited states ($\text{S}_1$) of both aggregates have non-vanishing TDMs.
Furthermore, the magnitudes of the TDMs are larger than those of the energetically higher-lying $\text{S}_2$ states.

The projections of the TDMs of the energetically lowest, emissive transitions of all three dimers onto both experimentally investigated crystal planes are depicted in Figure~\ref{fgr:5-tdm} c).
Note that we need to consider all symmetrically equivalent dimers. 
While the energy is degenerate for the different symmetric dimer configurations, their TDMs direction, and hence the polarization of the emitted light, differs.
We have taken this into account by the solid and dashed arrows representing the orientations of the TDMs in Figure~\ref{fgr:5-tdm}~c). 

Starting with the projection onto the $(223)$ facet in the left part of Figure~\ref{fgr:5-tdm}~c), the projections of the $\text{S}_1$ state TDM of Dimer 2 and the $\text{S}_3$ state TDM of Dimer 1 are both nearly parallel to the $(\bar{1}01)$ plane. The TDMs of the $\text{S}_1$ state of the different orientations of Dimer 3 on the other hand are oriented almost perpendicular to the plane.
For the the projection onto the $(\bar{1}01)$ facet, depicted in the right part of Figure~\ref{fgr:5-tdm}~c), the emissions of Dimer 1 and Dimer 3 are isotropic, arising from their different orientations in the crystal structure. Finally, the projection of the TDM of the $\text{S}_1$ state of Dimer 2 lies almost perfectly within the $(\bar{1}01)$ plane and is aligned perpendicular to the crystallographic $[010]$ direction. 

By repeating the TD-DFT calculations with the crystal structures at different temperatures, we can track changes in these excitonic couplings caused by the structures anisotropic variation with temperature.
As a result, drastic changes in the coupling can be excluded.
While the energies of the states shift slightly (see Figure~\ref{fgr:6-energies}), all dimers retain their respective H- or J-type couplings. 
Similarly, the TDM orientations and oscillator strengths show no significant changes either (\textit{c.f.}  Figure~S7 in SI).
Therefore, in addition to the absence of a discrete structural phase transition, as deduced from temperature dependent X-ray scattering (see SI~Section~S3), a change in the coupling regime of the neighboring monomers can also be excluded.  

\section{Discussion}

\begin{figure}[tb]
\centering
  \includegraphics[]{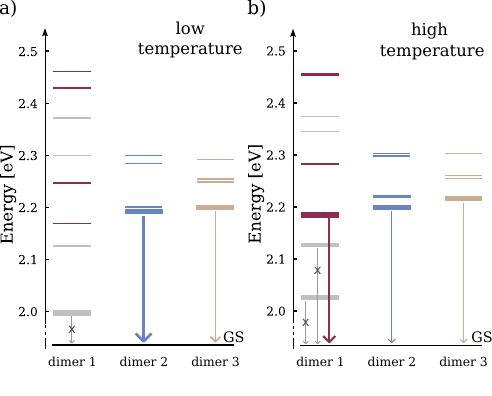}
  \caption{Energy level diagram of Dimers 1 (purple), 2 (blue) and 3 (yellow) at a) low and b) high temperatures resulting from TD-DFT calculations using the crystal structure determined at 83\,K and 293\,K, respectively. States with vanishing oscillator strength are depicted in gray. The occupation of states is illustrated schematically by thick lines and corresponding transitions to the ground state (GS) by vertical arrows.}
  \label{fgr:6-energies}
\end{figure}

The experimental photoluminescence data shows a drastic temperature-dependent change in the emission characteristics of $\upbeta$-ZnPc single crystals.
At high temperatures, the crystal exhibits a typical Frenkel-type but mostly unpolarized exciton emission. 
Upon cooling, there is almost complete loss of PL intensity, followed by the reemergence of a narrow, intense and highly polarized emission peak below 100\,K.
From the characterization of the ZnPc $\upbeta$-phase crystal structure as function of temperature, no structural phase transition can be identified well below the onset of the low-temperature emission.
Only a gradual, anisotropic temperature dependence of the crystallographic unit cell parameters is evident. 
Furthermore, TD-DFT calculations rule out the gradual change in the crystal structure as the reason for the change in emission.
These calculations show a gradual, but small shift in the energetic levels and TDMs of the calculated dimers.
However, there is no change in the coupling regime of the three investigated dimers, such as, for example, a change from H- to J-aggregate coupling due to a decrease in the intermolecular tilt below the value of the magic angle.\cite{Hestand.2018}
Figure~\ref{fgr:6-energies} schematically shows the thermal occupation and contributing emissive pathways for energy level diagrams calculated for the crystal structure at the highest and lowest accessible temperature of the X-ray diffraction experiments, 293\,K and 83\,K, respectively.

The drastic change in emission characteristics can be explained solely by the thermal population effects on different emissive states, as derived by TD-DFT.
The J-aggregate emission at low temperatures corresponds to the coupling within the $(010)$ plane, as described by Dimers 2 and 3.
Because the measured emission polarization is both perpendicular to the $[010]$ direction and oriented parallel to the $(\bar{1}01)$ plane, the dominant coupling can be further narrowed down to Dimer~2 along the $[101]$ direction.
Now, for increasing temperatures the thermal population of energetically higher states, such as $\text{S}_2$, increases while the population of the $\text{S}_1$ state decreases.
Since $\text{S}_1$ has a higher TDM than $\text{S}_2$, this ultimately results in a decrease of the overall emission intensity.
Correspondingly, the derived Boltzmann activation energy of the low temperature intensity decrease of $E_\textrm{A}=(9.5\pm2.1)$\,meV fits very well to the energy difference of Dimer 2's $\text{S}_1$ and $\text{S}_2$ states of 7.2\,meV.
At high temperatures, the H-aggregate-like coupled Dimer 1 contributes to the emission by population of its higher excited states, in particular its $\text{S}_3$ state. 
This process is mediated by scattering with high-momentum phonons, as derived from the steady-state emission being dependent on the thermal increase in phonon population following the Bose-Einstein statistic.

Furthermore, the energy difference between the J-type $\text{S}_1$ state of Dimer 2 and the energetically lower H-coupled $\text{S}_3$ state of Dimer 1 is 15\,meV.
This explains the blue shift of around 40\,meV observed in the emission spectra for decreasing temperatures.
These values fit well when considering that the experimental value is likely overestimated due to the effect of self-absorption as discussed above.

Due to its two different orientations in the crystal structure, the emission of Dimer 1 is isotropic with respect to the $(\bar{1}01)$ facet, which aligns with the experimental findings on the PL polarization at higher temperatures.
The microtome-cut $(223)$ facet also shows an overall decrease in polarization with rising temperatures, which can be explained by the thermal activation of additional emissive states included within the broad high-temperature emission. 

The time dependent analysis allows for an even more distinct attribution of the measured spectral features to different excitonic coupling directions in $\upbeta$-ZnPc.
The two main components at low temperatures can be attributed according to their polarization. 
The intense and fast component with a lifetime of 168\,ps is polarized parallel to the $(\bar{1}01)$ plane, corresponding to the $\text{S}_1$ state of Dimer 2.
The slower component with a lifetime of over 2\,ns is polarized perpendicular to this plane, corresponding to Dimer 3.
Due to the nearly vanishing energetic difference between the $\text{S}_1$ states of Dimers 2 and 3 at lower temperatures (see Figure~\ref{fgr:6-energies}), the two contributions cannot be distinguished based on their steady-state spectral signature alone. 

While this interpretation is consistent with our experimental and \textit{ab-initio} findings, it begs the question why the emission from the $(223)$ facet is so strongly polarized, when it should be dominated by emission from the S$_1$ states of Dimers 2 and 3 that have perpendicular TDMs yielding a totally unpolarized overall luminescence. 
While we consistently find contributions of Dimer 3 in the polarization dependent lifetime and steady-state analysis, its effective transition strength seems to be much weaker in the experiment than predicted by dimer calculations. 
We attribute this mismatch to an inherent restriction of the dimer model, as only short range coupling within small molecular units is considered. 


\section{Conclusion}

In this study, we demonstrated that, although it mimics a structural phase transition, the discontinuous change in fluorescence from J- to H-type in a single crystal of the $\upbeta$-phase of ZnPc is fully consistent with a thermalized dimer exciton model.
If thermal population of emissive and dark states is included in the analysis no substantial changes of the crystal structure are necessary to describe the observed luminescence data.
We thoroughly investigated temperature-dependent fluorescence via polarized steady-state and transient micro-PL measurements of different crystal facets accessible via microtome sectioning.
Complementing this analysis with a temperature-dependend structural analysis revealed no evidence of a structural phase transition.
Modeling the intermolecular coupling via TD-DFT calculations of dimers of neighboring molecules revealed different H- and J-aggregate coupling configurations.
After accounting for thermal occupation effects, the respective temperature dependence of these couplings fully explains the observed spectral changes.
Thus, this study serves as a representative case study to demonstrate, that observing an optical transition not necessarily indicates similar discrete changes in the underlying crystal structure.
The interplay and balance of a multitude of different coupling regimes in anisotropic low-symmetry crystal structures, which are common in crystalline molecular materials, can already result in a discontinuous luminescent behavior. 

\section{Methods}

\subsection*{Materials, crystal growth and sample preparation}
	
Single crystals of ZnPc were grown via horizontal vapor deposition (HVD) by sublimation of pre-purified powders.\cite{Laudise.1998,Jiang.2017} Sublimation took place at temperatures of about 750\,K and under a nitrogen flow of 40\,sccm. The as-grown crystals exhibit a needle-like shape with lengths up to 10\,mm and a rectangular cross-section of several microns in width (\textit{c.f.} Figure~S1 in SI).
 
For crystal sectioning by microtomy, ZnPc single crystals were embedded in an epoxy resin and then mounted on a Leica EM UC7 ultra-microtome. Cutting the crystal inside the transparent epoxy perpendicular to its long needle axis yields 100\,nm to 200\,nm thin cross-sectional slabs (\textit{c.f.} Figure~S1~b) in SI). 
After cutting, the sliced discs float on the surface of the adjacent water bath and each single crystalline disc was then picked up and transferred to TEM grids for optical characterization. 

\subsection*{X-ray diffraction studies}
	
Temperature dependent single crystal X-ray diffraction (XRD) studies in the range of 80\,K to 295\,K were conducted to confirm the $\upbeta$-phase growth of ZnPc single crystals and to investigate the anisotropy of its thermal expansion. 
The crystal data was collected on a Rigaku XtaLAB Synergy-R diffractometer with a HPA area detector and multi-layer mirror monochromated Cu$_{\text{K}\upalpha}$ radiation. 
The structures were solved using intrinsic phasing method,\cite{Sheldrick.2015} refined with the ShelXL program\cite{Sheldrick.2008} and expanded using Fourier techniques. 
All non-hydrogen atoms were refined anisotropically. 
The temperature dependent experiments were started at the lowest temperature and conducted with steps of 10\,K. 
After the target temperature was reached, a waiting period of five minutes before data collection allowed the temperature to stabilize.

According to the thermodynamically stable $\upbeta$-phase documented for ZnPc single crystals, the material crystallizes in the P$2_1$/n space group,\cite{Jiang.2017} fitting our findings. 
The derived unit cell parameters of ZnPc at 83(2)\,K are: $a=14.50898(12)$\,\AA, 	$b=4.84969(4)$\,\AA,  $c=17.12024(13)$\,\AA, $\alpha=\gamma=90$\,$^\circ$, $\beta=106.3465(8)$ and $V=1155.954(17)$\,\AA$^3$.  
A schematic of the molecular herringbone arrangement in the unit cell is shown in Figure~\ref{fgr:1-PL} b).

\subsection*{Transmission electron microscopy}

Selected area electron diffraction (SAED) measurements on microtome-cut ZnPc sections were conducted using a FEI Titan 80-300 equipped with a Gatan UltraScan 1000 camera and an incident electron beam of 300\,kV nominal to the sample surface. 
The resulting diffraction patterns were analyzed via a simulation and fitting routine using the \texttt{GARFIELD} software toolkit\cite{Marx.2025} resulting in a crystal plane with corresponding Miller indices of  approximately $h = 2$, $k = 2$ and $l = 3$ (\textit{c.f.} Figure~S4 in SI). 
This plane is tilted by 33\,$^\circ$ from the nominally cut $(010)$ plane perpendicular to the long needle axis aligned with the $[010]$ direction.

\subsection*{Micro photoluminescence}

All photoluminescence spectra were recorded with a micro-photoluminescence setup using a Princeton Instruments Acton SP2500i spectrometer together with a PIXIS 100 BReXelon CCD camera. 
For steady-state photoexcitation, a circularly polarized 685\,nm cw-laser was used and the polarization dependent photoluminescence detection was carried out by a combination of half-wave plate and linear polarizer. 
Time resolved emission spectroscopy (TRES) measurements were conducted via time correlated single photon counting (TCSP) using an Excelitas Technologies SPCM-AQRH-14 APD and a PicoHarp 300 TSCP module and excitation by a SuperK Fianium pulsed white light laser by NKT Photonics with a LLTF CONTRAST monochromator by Photon etc. set to 685\,nm. 
Spectral resolution was achieved by a GEMINI common-path Fourier interferometer by NIREOS. 
Global analysis of TRES data was performed using the \texttt{KiMoPack} python package\cite{Muller.2022} and final fitting of experimental life times was done within the \texttt{lmfit} framework.\cite{Newville.2025}

Both kinds of samples, the bulk single crystals as well as the microtome sectioned slabs on TEM-grids, were fixed on copper plates with conductive silver paint to ensure optimal thermal contact and placed in a contact cryostat offering a temperature range from room temperature down to 5\,K. 
Temperature dependent measurements were conducted by slowly heating the sample after cooling down to 5 K and ensuring thermal equilibrium by a sufficiently long delay at each temperature step. 
All emission spectra were transformed from the wavelength to the energy scale using the Jacobi-transformation.\cite{Mooney.2013} 

\subsection*{TD-DFT calculations}

Calculations of the excited state properties of different ZnPc dimers contained in the crystal were performed in the framework of time-dependent density functional theory (TD-DFT) utilizing the $\upomega$B97X-D3 functional\cite{Lin.2013} as implemented in the ORCA software package\cite{Neese.2020,Neese.2022} with the def2-SVP basis set.\cite{Weigend.2005}
The dimer structures for varying temperatures were obtained from the crystallographic data. 
\section*{Author Contributions}
J.P. and S.H. conceptualized the study.
L.S.-M. and S.H. grew the single crystals.
K.R. performed the X-ray structure analysis.
G.T., H.S. and L.S.-M. performed the ultra-microtomy.
L.S.-M. and K.S. performed the optical characterization, L.S.-M. analyzed the collected data. 
L.S.-M. visualized the research results
H.H.,and M.K. performed the SAED,  S.H. and L.S.-M. analyzed the collected data.
L.N.P. and R.M. conceptualized the TD-DFT calculations, L.N.P. and L.S.-M. performed the calculations and analyzed the data. 
L.S.-M., L.N.P. and S.H. wrote the original draft. 
All authors reviewed and edited the final draft. 
J.P., R.M., H.B., H.S., M.K. supervised and secured the funding.

\section*{Conflicts of interest}
There are no conflicts to declare.

\section*{Data availability}
The data supporting this article have been included as part of the Supplementary Information (SI).  
Any additional raw data required to reproduce the findings can be made available upon reasonable request. 

\section*{Acknoledgements}
L.S.-M. and S.H. acknowledge support by the Würzburg-Dresden Cluster of Excellence on Complexity and Topology in Quantum Matter ct.qmat (EXC 2147). 
L.N.P. acknoledges a fellowship from the FCI.
J.P. acknowledges financial support from the Bavarian State Ministry for Science and the Arts within the collaborative research network “Solar Technologies go Hybrid (SolTech)“.

\balance

\printbibliography
\end{document}


\maketitle

\begin{abstract}
\textit{$^{a}$~Experimental Physics VI, Julius-Maximilian University Würzburg, Am Hubland, 97074 Würzburg, Germany. E-mail: lisa.schraut-may@uni-wuerzburg.de, sebastian.hammer@uni-wuerzburg.de, jens.pflaum@uni-wuerzburg.de}
\textit{$^{b}$~Chair of Theoretical Chemistry, Julius-Maximilian University Würzburg, Am Hubland, 97074 Würzburg, Germany.}
\textit{$^{c}$~Institute of Inorganic Chemistry, Julius-Maximilian University Würzburg, Am Hubland, 97074 Würzburg, Germany.}
\textit{$^{d}$~Max Planck Institute for the Structure and Dynamics of Matter, Luruper Chaussee 149, 22761 Hamburg, Germany.}
\textit{$^{e}$~Institute of Physics and Röntgen Center for Complex Material Systems, Julius-Maximilian University Würzburg, Am Hubland, 97074 Würzburg, Germany.}
\textit{$^{f}$~Center for Applied Energy Research, 97074 Würzburg, Germany.}
\end{abstract}

\section{Microscopic images}

\begin{figure}[htb]
 \centering
 \includegraphics[width=0.9\linewidth]{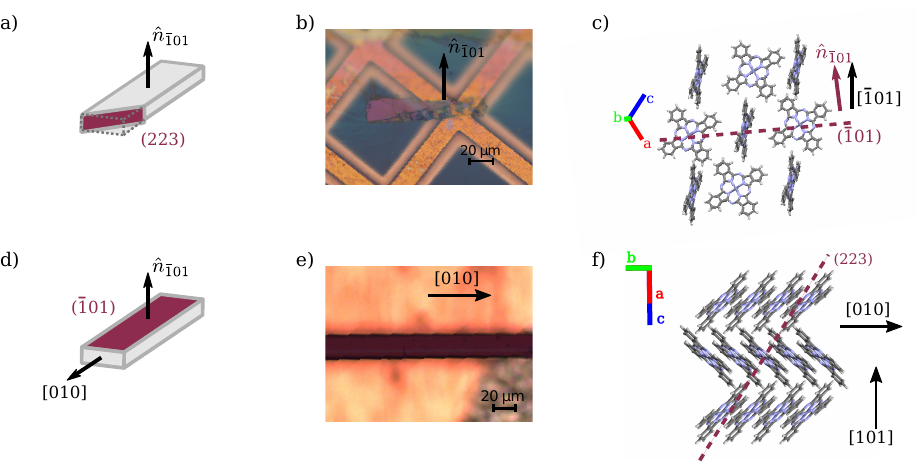}
 \caption{Model of a ZnPc single crystal illustrating the orientation of the probed a) $(223)$ and d) $(\bar{1}01)$ crystal facets. Microscopic images of b) a microtome cut $(223)$ slab on a TEM grid and e) a ZnPc single crystal with its natural $(\bar{1}01)$ facet facing out of the plane. Illustration of the alignment of the ZnPc monomers in the c) $(223)$ and f) $(\bar{1}01)$ planes, specific crystallographic directions are marked by rectangular parenthesis as well as planes by round parentheses. In c) furthermore the slight tilt between the crystallographic direction $[\bar{1}01]$ and the normal vector $\hat{n}_{\bar{1}01}$ of the $(\bar{1}01)$ plane due to the monoclinic symmetry of ZnPc is illustrated. }
 \label{fgr:1-Microscope}
\end{figure}

\newpage

\section{Self-absorption correction}

The photoluminescence of a an optically thick sample is subject to self-absorption before exiting its surface where it can be measured. Therefore, with knowledge of the samples absorption coefficient $\alpha (E)$, the detected emission $I_D(E)$ can be corrected to obtain the true intrinsic emission $I(E)$. According to Irkhin et. al\cite{Irkhin.2012}, the intrinsic photoluminescence of a thick (thicker than the absorption depth of the excitation) crystalline sample, can be calculated by
\begin{equation}
    I(E) = I_D(E) \left(1 + \frac{\alpha(E)}{\alpha_\text{exc}}\right)
\end{equation}
with the absorption coefficient of the excitation energy $\alpha_\text{exc}$. 

\begin{figure}[h!]
 \centering
 \includegraphics[width = 0.9\linewidth]{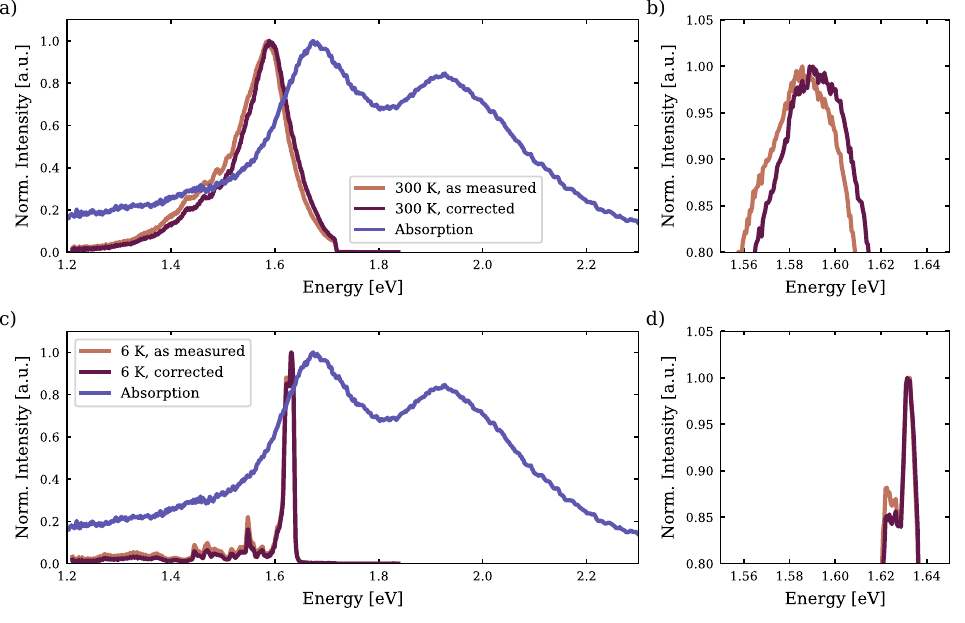}
 \caption{Self-Absorption correction. Comparison of the absorption, as-measured emission and self-absorption corrected emission at a) 300\,K and c) 6\,K. Close-ups at b) 300\,K and d) 6\,K emphasise the resulting shifts.}
 \label{fgr:3-Self-Absorption}
\end{figure}

The absorbance $A=x\alpha$ of the ZnPc $\upbeta$-phase sample of thickness $x$ was measured at room temperature on microtome cuts of the single crystals $(\bar{1}01)$-facet. Figure~\ref{fgr:3-Self-Absorption}~a) shows the absorption with the measured and corrected photoluminescence at room temperature. The correction leads to a shift of the spectrum by around 6\,meV towards higher energies (c.f. close-up in Figure~\ref{fgr:3-Self-Absorption}~b)). 

Due to the constraints of the used absorbance setup, no temperature dependent measurements were possible. Correction of the low temperature photoluminescence is hence only possible using the room temperature data and the results are therefore to be treated with caution. But, due to the narrowness of the low temperature PL signal, no significant shifts are expected. In Figure~\ref{fgr:3-Self-Absorption}~c) and d) the as-measured data at 6\,K is compared to the corrected spectrum, showing no obvious shifting. The hypsochromic shift of the emission with decreasing temperatures is therefore lowered from 40\,meV in the original data to only 34\,meV in the self-absorption corrected data. 

\newpage

\section{Temperature dependent structure characterization}

\subsection*{Crystal data for ZnPc}
\ce{C32H16N8O0Zn}, $M_r = 577.90$, orange needle, $0.566\times0.066\times0.025$\,\unit{\cubic\milli\meter}, monoclinic space group $P2_1/n$, 
$Z$ = 2, $r_\text{calcd}$ = 1.660\,\unit{\gram\cubic\centi\meter}, $m$ = 1.827\,\unit{\per\milli\meter}, $F(000)$ = 588, $R_1$ = 0.0456, $wR_2$ = 0.2553 at $T$ = 83(2)\,\unit{\kelvin}, 2297 independent reflections $[2q\leq148.474$°] and 82 parameters.

Reflections $(\bar{1} 0 1)$ and $(1 0 1)$ were omitted during refinement (partially covered with the beam-stop).

\begin{table}[h!]
\centering
\begin{tabular}{l|lllll}
Temperature / \unit{\kelvin} & $a$ / \unit{\angstrom} & $b$ / \unit{\angstrom} &$c$ / \unit{\angstrom} & $\beta$ / ° & $V$ / \unit{\cubic\angstrom} \\
\hline
83(2) & 14.50898(12) & 4.84969(4) & 17.12024(13) & 106.3465(8) & 1155.954(17) \\
93(2) & 14.51073(12) & 4.85017(4) & 17.12746(13) & 106.3358(8) & 1156.759(17) \\
103(2) & 14.51381(12) & 4.85081(4) & 17.13538(13) & 106.3260(9) & 1157.750(17) \\
113(2) & 14.51586(12) & 4.85142(4) & 17.14141(13) & 106.3100(9) & 1158.562(17) \\
123(2) & 14.51889(12) & 4.85209(4) & 17.14777(13) & 106.2990(9) & 1159.459(17) \\
133(2) & 14.52180(13) & 4.85279(4) & 17.15506(14) & 106.2849(9) & 1160.435(18) \\
143(2) & 14.52387(12) & 4.85354(4) & 17.16284(13) & 106.2697(9) & 1161.396(17) \\
153(2) & 14.52650(13) & 4.85424(4) & 17.16965(14) & 106.2549(9) & 1162.322(18) \\
163(2) & 14.52928(13) & 4.85509(4) & 17.17776(13) & 106.2432(9) & 1163.367(17) \\
173(2) & 14.53224(13) & 4.85617(4) & 17.18451(13) & 106.2300(9) & 1164.398(17) \\
183(2) & 14.53477(12) & 4.85704(4) & 17.19256(13) & 106.2177(8) & 1165.428(17) \\
193(2) & 14.53718(12) & 4.85814(4) & 17.19988(13) & 106.2004(8) & 1166.484(17) \\
203(2) & 14.54013(12) & 4.85933(4) & 17.20743(13) & 106.1804(8) & 1167.637(17) \\
213(2) & 14.54315(12) & 4.86066(4) & 17.21518(13) & 106.1626(8) & 1168.830(17) \\
223(2) & 14.54596(13) & 4.86236(4) & 17.22248(13) & 106.1379(8) & 1170.108(17) \\
233(2) & 14.54893(13) & 4.86404(4) & 17.23082(14) & 106.1167(9) & 1171.443(18) \\
243(2) & 14.55220(13) & 4.86548(4) & 17.23899(14) & 106.0979(9) & 1172.720(18) \\
253(2) & 14.55553(14) & 4.86697(4) & 17.24760(15) & 106.0740(9) & 1174.075(19) \\
263(2) & 14.55868(13) & 4.86846(4) & 17.25549(14) & 106.0501(9) & 1175.367(18) \\
273(2) & 14.56209(13) & 4.86958(4) & 17.26315(14) & 106.0298(9) & 1176.554(18) \\
283(2) & 14.56579(13) & 4.87068(4) & 17.27173(14) & 106.0101(9) & 1177.821(18) \\
293(2) & 14.56912(13) & 4.87187(4) & 17.28061(14) & 105.9898(9) & 1179.103(18) \\
\end{tabular}
\caption{Temperature dependent crystallographic data of ZnPc, the space group is $P2_1/n$ for all temperatures with $\alpha$ = $\gamma$ = 90°. }
\label{tbl:crystal-para}
\end{table}

\begin{figure}[h!]
 \centering
 \includegraphics[width=\linewidth]{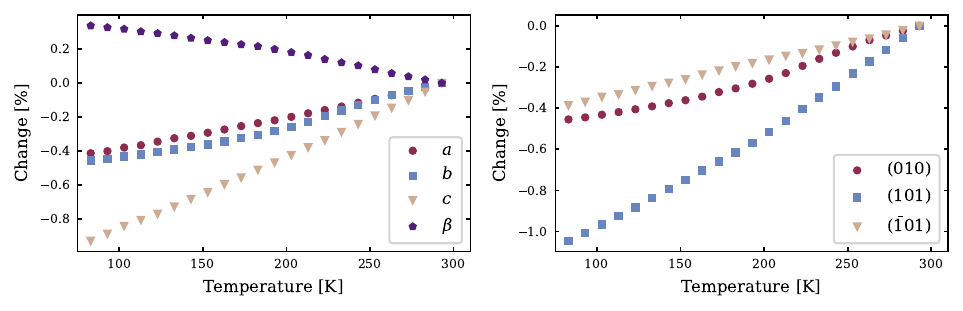}
 \caption{Change of unit cell parameters $a$, $b$, $c$ and $\beta$ (left panel) and spacing of $(010)$, $(101)$ and $(\bar{1}01)$-planes (right panel) with temperature. The parameters are normalized to their respective room temperature values. }
 \label{fgr:Crystal-Structure}
\end{figure}

\newpage
\newpage

\section{Transmission electron microscopy}

\begin{figure}[htb]
 \centering
 \includegraphics[width = 0.8\linewidth]{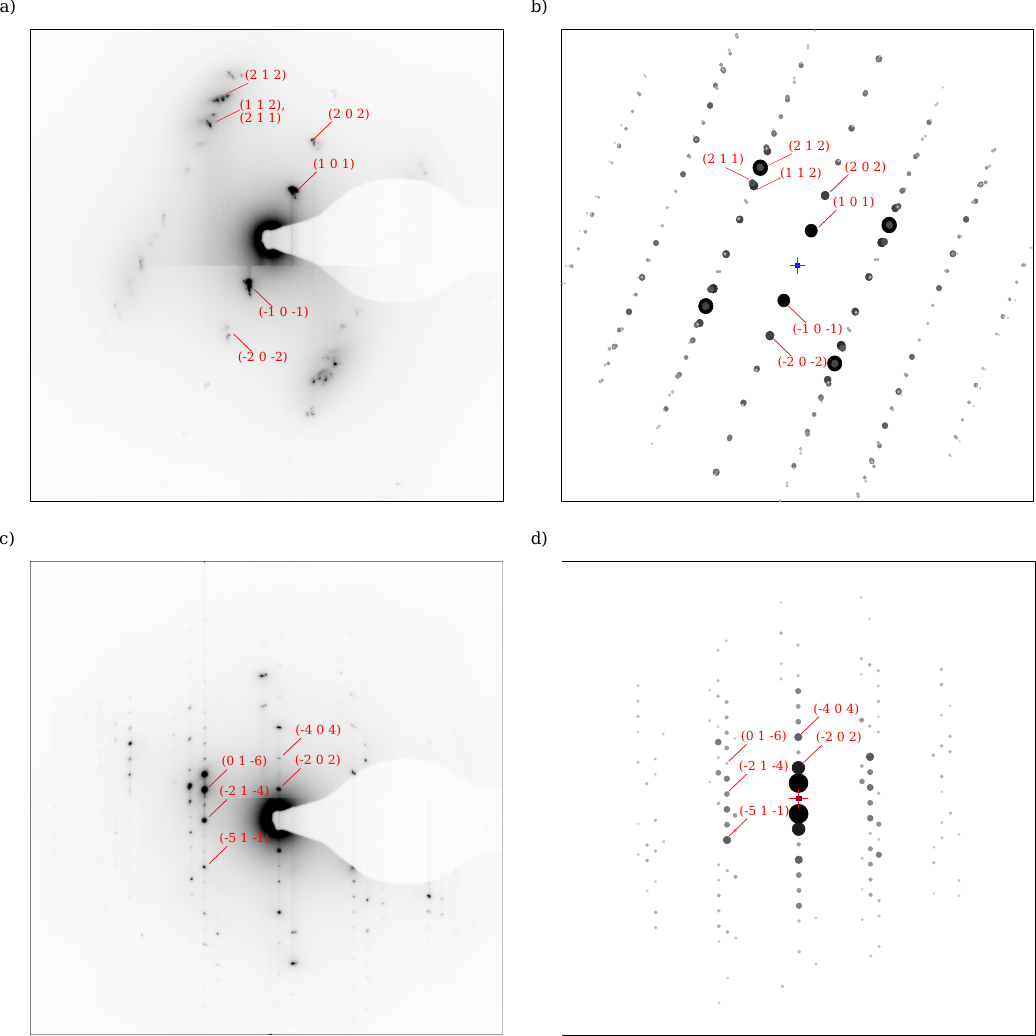}
 \caption{Comparison of the measured selected area electron diffraction (SAED) images of microtome cuts on TEM grids of the the a) naturally occuring upper facet and b) cross-sections of the ZnPc $\upbeta$-phase. Corresponding simulated TEM patterns of the b) $(\bar{1}01)$ and d) $(223)$ plane created by the software toolkit \texttt{GARFIELD}\cite{Marx.2025}. For the simulations a isotropic mosaicity of the sample of b) 5.5° and d) 2° was estimated.}
 \label{fgr:tem}
\end{figure}

\section{Excimer emission}

\begin{figure}[htb!]
 \centering
 \includegraphics[]{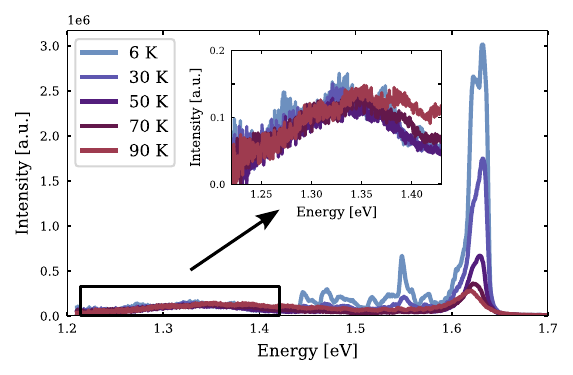}
 \caption{Low temperature emission of the $(\bar{1}01)$ facet. The inset shows the broad, low-intensity emission around 1.35\,eV fitting to the excimer emission of the ZnPc $\upalpha$-phase.}
 \label{fgr:4-Excimer}
\end{figure}

\section{Lifetime analysis}

\subsection{Global analysis}

\begin{figure}[h!]
 \centering
 \includegraphics[width = \linewidth]{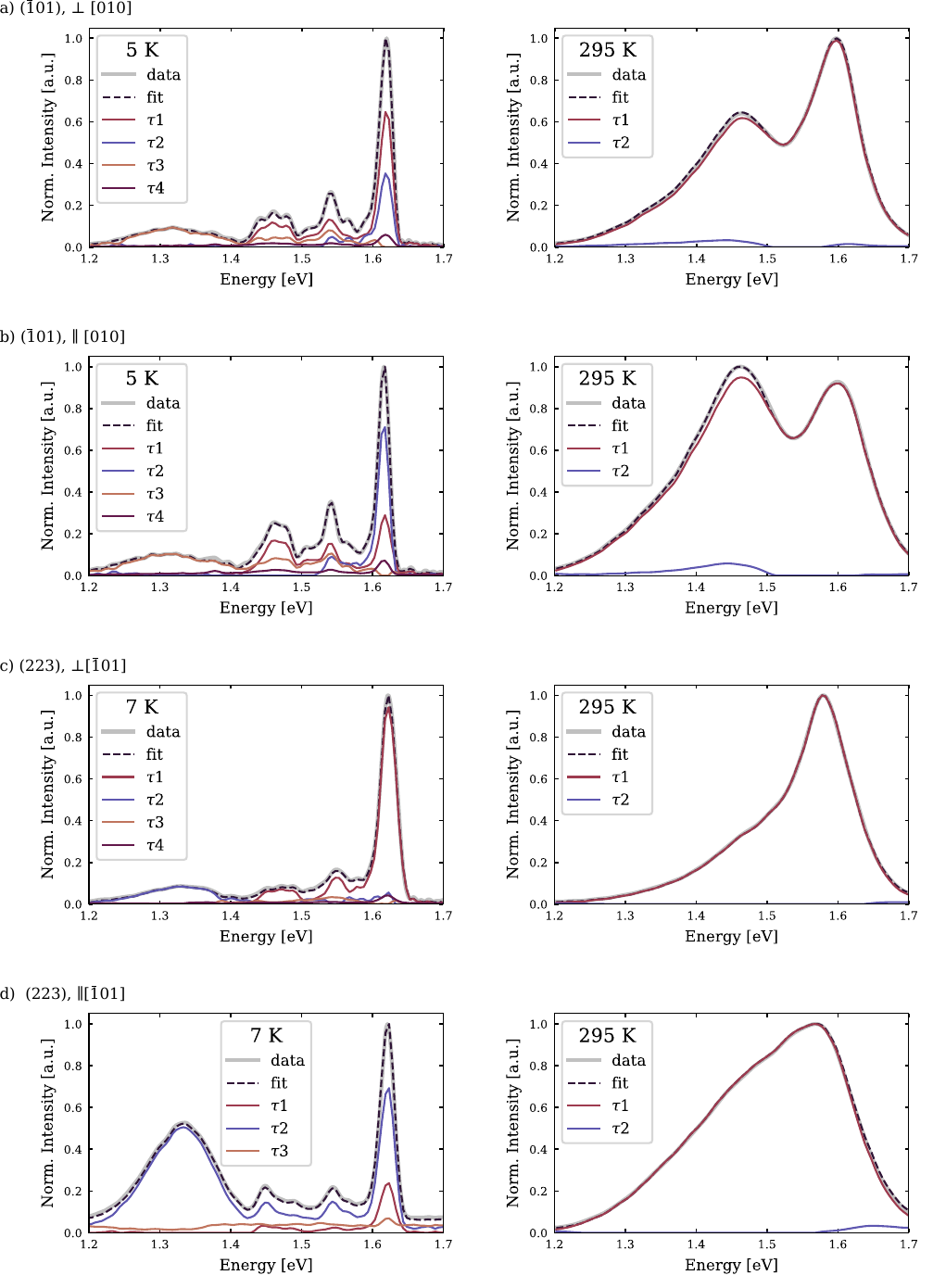}
 \caption{Global analysis of the TRES data at highest and lowest measured temperature for each measured facet and polarization direction: the $(\bar{1}01)$ facet with polarization a) perpendicular and b) along the $[010]$ direction and the $(223)$ facet with polarization c) perpendicular and d) along the $[\bar{1}01]$ direction. The data analysis was done using the \texttt{KiMoPack} python package\cite{Muller.2022}.}
 \label{fgr:GA}
\end{figure}

The time-dependent emission spectroscopy (TRES) data was analyzed by global analysis (GA) using the \texttt{python} package \texttt{KiMoPack}\cite{Muller.2022}. First, a singular value decomposition (SVD) of the data was performed to derive the number of time dependent components comprising the data. Then, a GA using two to four indepent exponentially decaying components convoluted with the experimentally derived instrument response function (IRF) of the setup was conducted. The resulting decay associated spectra (DAS) integrated over time for comparison with the measured integrated spectra are shown in Figure~\ref{fgr:GA}. 

Hereby, the different contributions to the spectrum can be disentangled. Firstly, all datasets contain a comparatively slow (several ns), low intensity and spectrally broad contribution. Due to its unspecific characteristics, this contribution cannot be appointed to a specific emissive state and is rather attributed to a multi-component background signal. Besides this contribution, the spectra at room temperature are fully represented by a single component spectrally fitting to a Frenkel-type excitonic emission with a accompanying vibronic progression. Towards lower temperatures, the GA reveals two to three contributions additional to the before mentioned broad one. The broad emission at 1.35\,\unit{\electronvolt} can clearly be distinguished from the narrow again Frenkel-type emission at 1.6\,\unit{\electronvolt}  by its different emission lifetime. This main peak at 1.6\,\unit{\electronvolt} is at temperatures below 100\,K comprised of two components, whose spectral shapes are similar, but differing in both emission intensity and lifetime. 

\clearpage

\subsection{Multi-exponential fitting}

For a sole analysis of the main spectral component, a separate multi-exponential non-linear least-square fitting routine using the \texttt{lmfit} framework\cite{Newville.2025} is employed. The TRES data around 1.6\,\unit{\electronvolt} is integrated and the resulting time dependent emission decay fitted by a bi- or tri-exponential ($n=2,3$) decay convoluted with the setups IRF:
\begin{equation}
    I(t) = \sum_{i=1}^{n}A_\textrm{i}\exp\left(-\frac{t}{\tau_\textrm{i}}\right)\times IRF(t) \qquad .
\end{equation}
The resulting fit parameters for the probed $(\bar{1}01)$ and $(223)$ facets and both polarization directions, respectively, are listed in Tables~\ref{tbl:lifetime-101} and \ref{tbl:lifetime-223}. 

\begin{table}[ht]
    \centering
    \begin{subtable}{0.85\textwidth}
        \centering
        \caption{$\perp [010]$}
        \begin{tabular}{l||ll|ll|ll}
        T / \unit{\kelvin} & $\tau_1$ / \unit{\pico\second} &$A_1$ / \unit{cps} & $\tau_2$ / \unit{\nano\second} &$A_2$ / \unit{cps}& $\tau_3$ / \unit{\nano\second} &$A_3$ / \unit{cps} \\
        \hline
        5 & $167\pm8$ & $1686\pm1$ & $2.14\pm0.01$ & $23.6\pm0.1$ & $11.9\pm0.1$ & $1.97\pm0.03$ \\
        80 & $227\pm8$ & $696.1\pm0.8$ & $2.21\pm0.09$ & $2.7\pm0.1$ & $21.5\pm0.3$ & $1.39\pm0.02$ \\
        200 & $251\pm8$ & $1394\pm2$ & - & - & $23.3\pm0.2$ & $2.80\pm0.02$ \\
        295 & $245\pm8$ & $1607\pm1$ & - & - & $29.8\pm0.3$ & $1.92\pm0.01$ \\
        \end{tabular}
    \end{subtable}

    \begin{subtable}{0.85\linewidth}
        \centering
        \caption{$\parallel [010]$}
        \begin{tabular}{l||ll|ll|ll}
        T / \unit{\kelvin} & $\tau_1$ / \unit{\pico\second} &$A_1$ / \unit{cps} & $\tau_2$ / \unit{\nano\second} &$A_2$ / \unit{cps}& $\tau_3$ / \unit{\nano\second} &$A_3$ / \unit{cps} \\
        \hline
        5 & $210\pm8$ & $610.4.0\pm0.5$ & $1.55\pm0.01$ & $15.8\pm0.2$ & $8.58\pm0.08$ & $1.47\pm0.02$ \\
        80* & $228\pm8$ & $231\pm3$ & $0.69\pm0.03$ & $30\pm3$ & $24.8\pm0.5$ & $2.13\pm0.02$ \\
        200 & $258\pm8$ & $563.7\pm0.8$ & - & - & $22.7\pm0.3$ & $1.08\pm0.01$ \\
        295 & $245\pm8$ & $706.0\pm0.7$ & - & - & $28.5\pm0.4$ & $0.77\pm0.01$ \\
        \end{tabular}
    \end{subtable}

    \caption{Multi-exponential fit results for the $(\bar{1}01)$ facet with polarization a) perpendicular and b) parallel to the $[010]$ direction. *Data was measured with a shorter acquisition time and therefore has to be treated with caution.}
    \label{tbl:lifetime-101}
\end{table}

\begin{table}[ht]
    \centering
    \begin{subtable}{0.85\textwidth}
        \centering
        \caption{$\perp [\bar{1}01]$}
        \begin{tabular}{l||ll|ll|ll}
        T / \unit{\kelvin} & $\tau_1$ / \unit{\pico\second} &$A_1$ / \unit{cps} & $\tau_2$ / \unit{\nano\second} &$A_2$ / \unit{cps}& $\tau_3$ / \unit{\nano\second} &$A_3$ / \unit{cps} \\
        \hline
    7 & $283\pm8$ & $521.1\pm0.4$ & $2.08\pm0.03$ & $6.23\pm0.09$ & $15.4\pm0.2$ & $0.79\pm0.01$ \\
    80 & $328\pm8$ & $378.0\pm0.5$ & $2.27\pm0.03$ & $10.1\pm0.2$ & $18.9\pm0.3$ & $0.99\pm0.02$ \\
    200 & $303\pm8$ & $2034\pm2$ & - & - & $9.26\pm0.05$ & $11.14\pm0.07$ \\
    295 & $360\pm8$ & $2612\pm3$ & - & - & $14.5\pm0.3$ & $0.029\pm0.001$ \\
        \end{tabular}
    \end{subtable}

    \begin{subtable}{0.85\linewidth}
        \centering
        \caption{$\parallel [\bar{1}01]$}
        \begin{tabular}{l||ll|ll|ll}
        T / \unit{\kelvin} & $\tau_1$ / \unit{\pico\second} &$A_1$ / \unit{cps} & $\tau_2$ / \unit{\nano\second} &$A_2$ / \unit{cps}& $\tau_3$ / \unit{\nano\second} &$A_3$ / \unit{cps} \\
        \hline
    7* & $370\pm8$ & $175.0\pm0.8$ & $1.16\pm0.02$ & $21.2\pm0.9$ & $15.6\pm0.2$ & $1.18\pm0.02$ \\
    80 & $340\pm8$ & $225.5\pm0.5$ & $2.00\pm0.03$ & $9.4\pm0.2$ & $22.7\pm0.3$ & $1.05\pm0.01$ \\
    200 & $309\pm8$ & $1317\pm1$ & - & - & $9.26\pm0.05$ & $7.14\pm0.05$ \\
    295 & $376\pm8$ & $1600\pm2$ & - & - & $10.9\pm0.2$ & $1.53\pm0.03$ \\
        \end{tabular}
    \end{subtable}

    \caption{Multi-exponential fit results for the $(223)$ facet with polarization a) perpendicular and b) parallel to the $[\bar{1}01]$ direction. *Data was measured with a shorter acquisition time and therefore has to be treated with caution.}
    \label{tbl:lifetime-223}
\end{table}

\newpage

\section{DFT calculations}

\begin{table}[h!]
\centering
\begin{tabular}{l||lllll|lllll}
 & 83\,K& & & & & 293\,K& & & &\\
 Dimer 1 & E / \unit{\electronvolt} & $f_\textrm{osc}$ & $\mu_x$& $\mu_y$& $\mu_z$& E / \unit{\electronvolt} & $f_\textrm{osc}$ & $\mu_x$& $\mu_y$&$\mu_z$\\
\hline
$\text{S}_1$ & 1.997& 0 & 0 & 0 & 0  & 2.026 & 0& 0& 0&0\\
$\text{S}_2$ & 2.126& 0 & 0 & 0 & 0  & 2.127 & 0& 0& 0&0\\
$\text{S}_3$ & 2.169& 0.705& 1.332& 2.761& 1.966& 2.184& 0.749& 1.640& 2.852&1.780\\
$\text{S}_4$ & 2.247& 0.428& 2.179& 0.758& -1.563& 2.283& 0.472& 2.105& 0.388&-1.962\\
$\text{S}_5$ & 2.299& 0 & 0 & 0 & 0  & 2.345& 0& 0& 0&0\\
$\text{S}_6$ & 2.372& 0 & 0 & 0 & 0  & 2.374& 0& 0& 0&0\\ 
$\text{S}_7$ & 2.430& 0.658& -1.927& 0.316& 2.691& 2.453& 0.545& -2.184& -0.386&2.036\\
$\text{S}_8$ & 2.461& 0.365& 1.470& 1.800& 0.805& 2.455& 0.391& 0.710& 1.775&1.686\\
\end{tabular}

\bigskip

\begin{tabular}{l||lllll|lllll}
 & 83\,K& & & & & 293\,K& & & &\\
 Dimer 2 & E / \unit{\electronvolt} & $f_\textrm{osc}$ & $\mu_x$& $\mu_y$& $\mu_z$& E / \unit{\electronvolt} & $f_\textrm{osc}$ & $\mu_x$& $\mu_y$&$\mu_z$\\
\hline
$\text{S}_1$ & 2.194& 0.778&  1.722& -0.073& 3.391& 2.199 & 0.798 & 2.109& -0.026&3.221\\
$\text{S}_2$ & 2.201& 0.053& 0.448& 0.389& 0.789& 2.219 & 0.030 & 0.176& 0.688&0.197\\
$\text{S}_3$ & 2.284& 1.179& -3.613& 0.166& 2.826& 2.298 & 1.173 & -3.376& 0.645&3.004\\
$\text{S}_4$ & 2.300& 0.641& 0.125& 3.363& -0.230& 2.303 & 0.648 & 0.607& 3.264&0.682\\
\end{tabular}

\bigskip

\begin{tabular}{l||lllll|lllll}
 & 83\,K& & & & &    293\,K&&&&\\
 Dimer 3 & E / \unit{\electronvolt} & $f_\textrm{osc}$ & $\mu_x$& $\mu_y$& $\mu_z$&    E / \unit{\electronvolt} &$f_\textrm{osc}$ &$\mu_x$&$\mu_y$&$\mu_z$\\
\hline
$\text{S}_1$ & 2.200& 1.505& 3.651& 1.118& -3.652&    2.216 &1.410 &3.808&1.353&-3.106\\
$\text{S}_2$ & 2.249& 0.559& 0.917& 1.751& 2.498&    2.255 &0.647 &-1.982&1.349&-2.444\\
$\text{S}_3$ & 2.255& 0.657& 2.221& -2.486& 0.889&    2.261 &0.752 &-0.729&2.910&2.140\\
$\text{S}_4$ & 2.292& 0.179& 0.313& 1.574& 0.787&    2.303 &0.078 &0.293&0.929&0.663\\
\end{tabular}

\caption{Excitation energies $E$, oscillator strength $f_\textrm{osc}$ and transition dipole moments $\mathbf{\mu}$ of the energetically lowest excited states of the ZnPc Dimers 1, 2 and 3 in Cartesian coordinates with unit vectors $\mathbf{x}$ and $\mathbf{y}$ oriented along crystallographic unit cell vectors $\mathbf{a}$ and $\mathbf{b}$, respectively. The shown values were calculated using the crystal structures at 293\,K and 83\,K.}
\label{tbl:td-dft}
\end{table}

\begin{figure}[htb!]
 \centering
 \includegraphics[width = 1\linewidth]{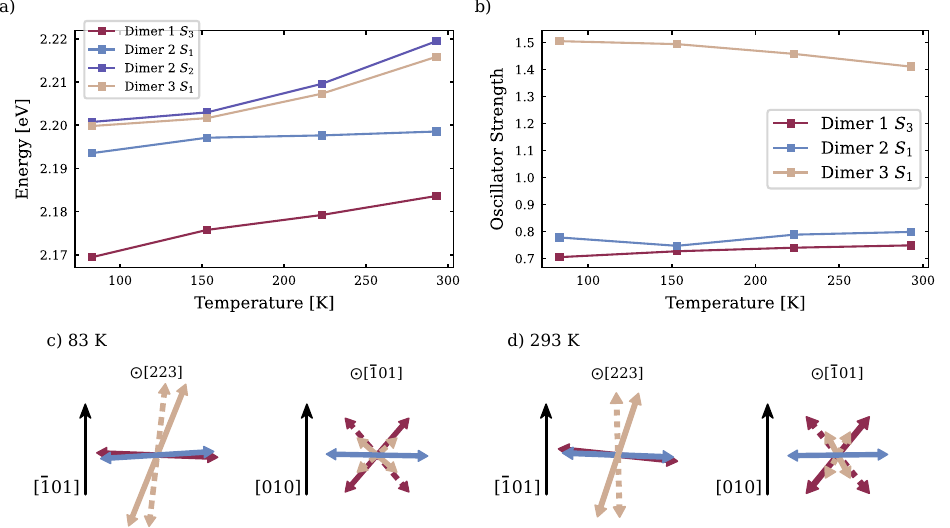}
 \caption{Temperature dependence of the a) excitation energies and b) oscillator strengths of the lowest bright states derived by TD-DFT. Projections of the calculated transition dipole moments of the energetically lowest bright transitions at c) 83\,K and d) 293\,K onto the $(223)$ (left) and $(\bar{1}01)$ plane (right).}
 \label{fgr:tddft}
\end{figure}

\newpage

\bibliographystyle{plain}
\bibliography{SI-bib}